\documentclass[a4paper]{jpconf}
\usepackage{graphicx}
\begin{document}
\title{Stable beam experiments in wide energy ranges serving low energy nuclear astrophysics}

\author{Gy. Gy\"urky}

\address{Institute for Nuclear Research (Atomki), H-4001 Debrecen, Hungary}

\ead{gyurky@atomki.mta.hu}

\begin{abstract}

In experimental nuclear astrophysics it is common knowledge that reaction cross sections must be measured in the astrophysically relevant, low energy ranges or at least as close to them as possible. In most of the cases, however, it is impossible to reach such low energies. The reactions must therefore be studied at higher energies and the cross sections must be extrapolated to lower ones. In this paper the importance of cross section measurements in wide energy ranges are emphasized and a few examples are shown from the areas of hydrogen burning processes and heavy element nucleosynthesis.

\end{abstract}

\section{Introduction}

For the modeling of astrophysical phenomena, including stellar evolution itself and various nucleosynthetic processes, the rates of nuclear reactions participating in a given process must be known at the corresponding stellar temperatures. The reaction rates can be obtained from the cross sections in the relevant energy ranges, i.e. in their Gamow windows. In most of the astrophysical processes the temperatures are so low that the Gamow windows are located way below the Coulomb barrier and thus the cross sections are tiny. It happens actually very rarely that a reaction can be studied experimentally within its whole Gamow window.

Table \ref{table:reactions} shows a few examples of the location of the Gamow windows and the corresponding cross sections in the case of some reactions mentioned in this paper. As one can see, the cross sections are indeed very low preventing the direct measurements in the Gamow windows. It is worth noting that the LUNA collaboration exploiting the world unique deep underground accelerator has pushed down the limit of measurable cross sections and successfully studied some reactions in their Gamow windows \cite[and references therein]{BROGGINI201855}.

If no experimental cross section data are available in the Gamow window, the reaction rate calculations must rely on theoretical cross sections. The reliability of these theoretical cross sections depends strongly on the available data at higher energies. In various hydrogen burning processes low mass nuclei are involved which are characterized by low level densities. In these cases the R-matrix approach is - among others - a standard procedure for obtaining low energy cross sections. For a robust R-matrix extrapolation, higher energy experimental data in a wide energy range is necessary as information is needed about the direct capture component as well as on some wide resonances which may influence the low energy extrapolation. If we go to the heavier mass regions, as for example in the cases of nucleosynthesis processes above iron, the nuclear level density is strongly increasing. Here the Hauser-Feshbach statistical model can provide the cross sections. This model uses several nuclear input parameters which can be tested at higher energies where the cross sections are large enough. If the models are tuned at higher energies, the calculation at low, astrophysical energies may also become more reliable.

\begin{center}
\begin{table}[th]
\caption{\label{table:reactions} Relevant temperatures$^*$, Gamow windows and the corresponding cross sections of some selected nuclear reactions discussed in this paper}
\centering
\begin{tabular}{lccc}
\br
Reaction & Temperature [10$^6$\,K] & Gamow window [keV] & cross section [barn]\\
\mr
$^3$He($\alpha,\gamma$)$^7$Be & 15 & 15\,--\,30 & $<$ 10$^{-20}$\\
$^{14}$N(p,$\gamma$)$^{15}$O & 15 & 20\,--\,35 & $<$ 10$^{-18}$\\
$^{17}$O(p,$\gamma$)$^{18}$F & 30 & 35\,--\,60 & $<$ 10$^{-15}$\\
$^{191}$Ir($\alpha,\gamma$)$^{195}$Au & 2000 & 6800\,--\,9700 & $<$ 10$^{-10}$\\
$^{197}$Au($\alpha,\gamma$)$^{201}$Tl & 2000 & 7000\,--\,9700 & $<$ 10$^{-10}$\\
\br
\end{tabular}\\
\flushleft\footnotesize $^*$ For the first three hydrogen burning reactions the quoted temperatures correspond to the lowest relevant ones. In the Gamow windows for higher temperatures, the cross sections are significantly higher and direct experimental data may exist.
\end{table}
\end{center}

\section{The activation method}

In this paper a few examples will be shown briefly where experimental data at energies much higher than the Gamow windows are useful and were obtained recently at Atomki. All these reactions were studied using the activation method, i.e. the cross sections were determined by the decay measurement of the created radioactive isotopes. Naturally, the activation method can be applied only in those cases where the reaction product is radioactive (and possesses some measurable decay signature), but in these cases the method has some clear advantages. The beam induced, or environmental background is typically less problematic in the case of an activation experiment. The decay occurs isotropically, hence there is no need to take into account the often problematic angular distribution effects. The activation method provides directly the astrophysically relevant total cross section and the results are largely independent from the data obtained from in-beam experiments. Therefore, the activation technique is a very useful complementary method or often - like in the case of heavy element nucleosynthesis processes - the only way to obtain reaction cross sections. Further details of the activation method in nuclear astrophysics related experiments can be found in a recent review \cite{Gyurky2019}. 

\section{Some selected reactions studied recently}
\subsection{The $^{17}$O(p,$\gamma$)$^{18}$F reaction}

The $^{17}$O(p,$\gamma$)$^{18}$F reaction plays an important role in advanced hydrogen burning processes in various stellar conditions from red giants to nova explosions. Recent experiments concentrated on the measurement of its low energy cross section \cite{PhysRevLett.109.202501}, while at higher energies and in a wide energy range only two, somewhat contradicting data sets are available \cite{ROLFS197329,PhysRevC.86.055801}. As $^{18}$F is radioactive and decays by positron emission, the activation method is applicable, however, it was used for this reaction only at low energies. In our work we measured the cross section by activation between 500\,keV and 1.8\,MeV. Our results indicated e.g. that the cross sections measured by C. Rolfs \cite{ROLFS197329} are too high and may contain some experimental error. Our largely independent activation cross sections were used to carry out a new R-matrix calculation. Moreover, at 500\,keV the available cross sections could be compared and some contradictions were identified. For further details see \cite{PhysRevC.95.035805}.

\subsection{The $^3$He($\alpha,\gamma$)$^7$Be reaction}

The $^3$He($\alpha,\gamma$)$^7$Be reaction is crucial for both the solar hydrogen burning processes and the primordial nucleosynthesis in the big bang. Many experiments were devoted to the study of this highly important reaction in recent years, but the precision of the cross section is still not at the required level for the astrophysical models \cite[and references therein]{BORDEANU20131}. The available experimental data extend up to about 3\,MeV, but the higher energy range is completely unexplored although R-matrix extrapolations may also be sensitive to even higher energy range. Moreover, recently the observation of a new resonance was claimed in the $^{6}$Li(p,$\gamma$)$^{7}$Be reaction, which leads to the same final nucleus \cite{HE2013287}. The existence of this resonance may also influence the description of the $^3$He($\alpha,\gamma$)$^7$Be reaction. We have studied the $^3$He($\alpha,\gamma$)$^7$Be reaction using the activation method in the energy range corresponding to this putative resonance. No resonance was observed and the energy dependence of the cross section was found to be somewhat different from what the R-matrix calculations show. Details of the experiment and the results can be found in \cite{PhysRevC.99.055804}. Further measurements connecting this energy range to the available data at lower energies are in preparation.

\subsection{The $^{14}$N(p,$\gamma$)$^{15}$O reaction}

The $^{14}$N(p,$\gamma$)$^{15}$O reaction is another key reaction of nuclear astrophysics as this reaction determines the rate of the CNO cycle hydrogen burning in various astrophysical sites. In spite of the tremendous experimental effort devoted to this reaction, the knowledge of its reaction rate still remains poor compared to the requirements of astrophysical models motivated by high precision observations \cite{doi:10.1146/annurev.nucl.012809.104505}. Owing probably to the short half-life of $^{15}$O (t$_{1/2}$\,=\,2.0\,min), no modern measurement of the $^{14}$N(p,$\gamma$)$^{15}$O has been carried with the activation technique. Therefore, the aim of our ongoing project at Atomki is to provide activation cross section data which are in many aspects independent from the in-beam $\gamma$-spectroscopy data sets. The activation data may be used to check the reliability of the available data sets and eventually to increase the precision of the  $^{14}$N(p,$\gamma$)$^{15}$O reaction rate.

\begin{figure}[th]
\vspace{1mm}
\includegraphics[angle=270,width=0.5\textwidth]{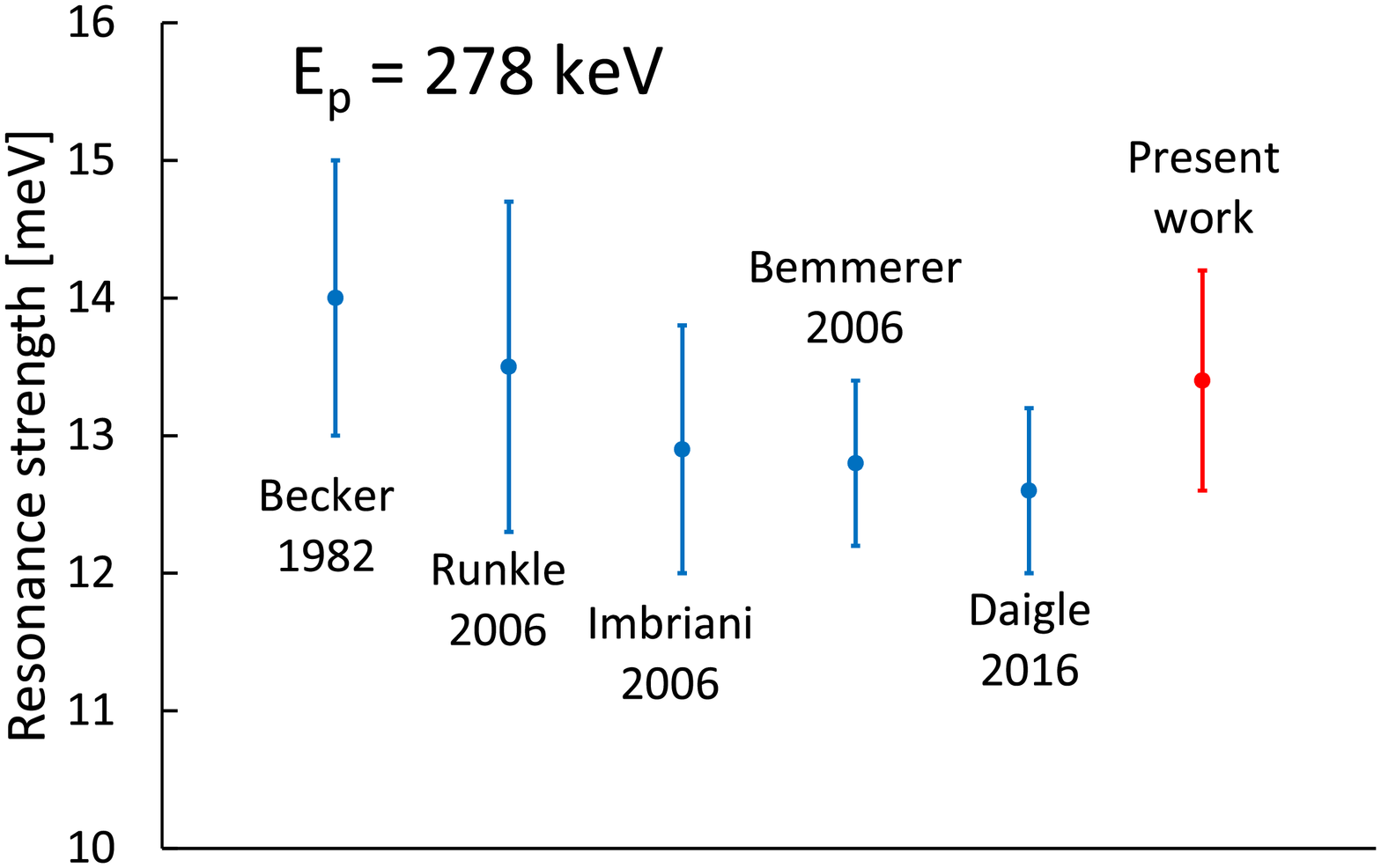}
\includegraphics[angle=270,width=0.6\textwidth]{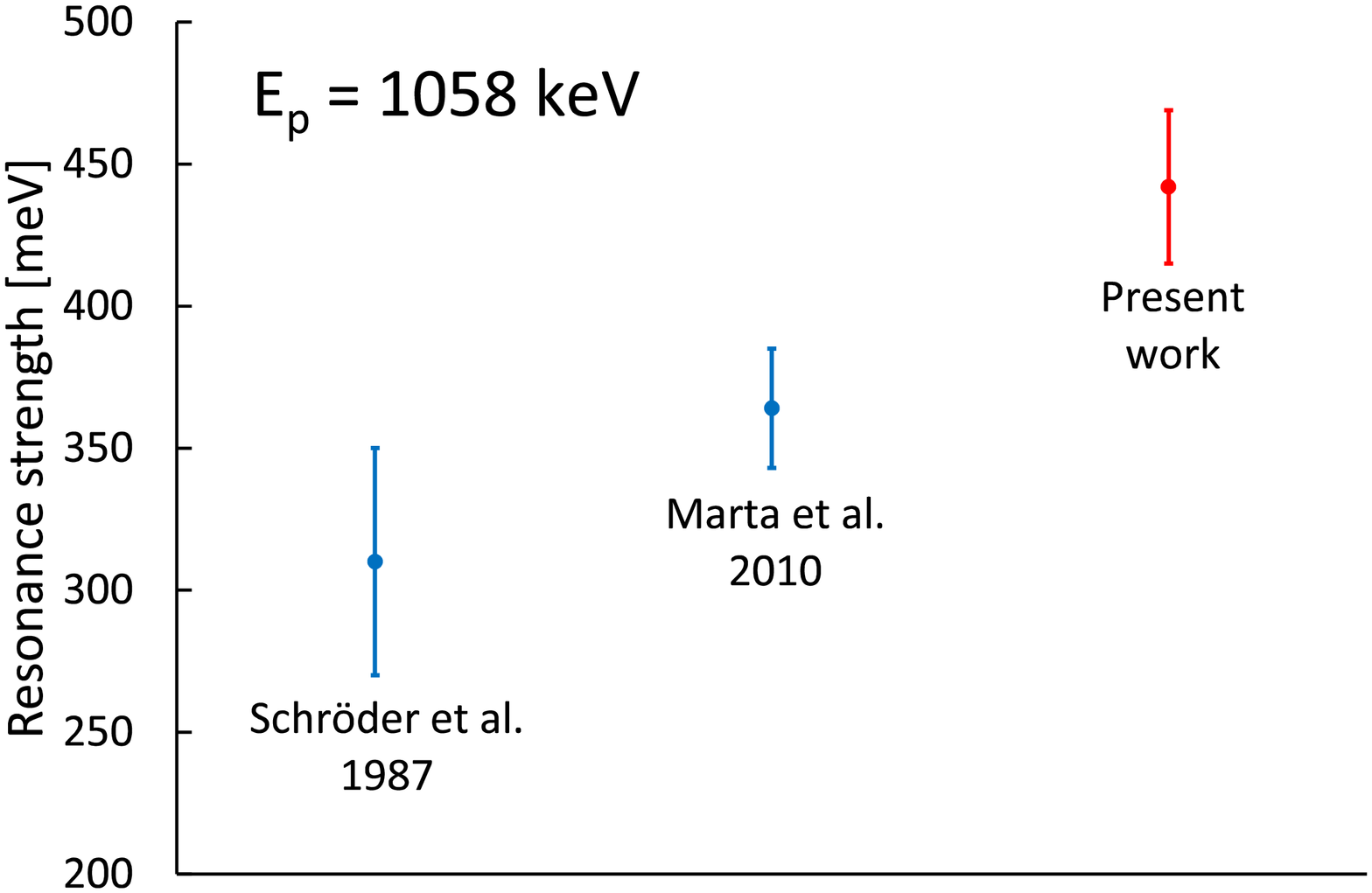} \\
\caption{\label{fig:strengths}Strengths of the two measured resonances in $^{14}$N(p,$\gamma$)$^{15}$O from the literature and from the present work. For the references see the bibliography in \cite{PhysRevC.100.015805}.}
\end{figure}

The strengths of two strong narrow resonances in $^{14}$N(p,$\gamma$)$^{15}$O have already been measured and the results published \cite{PhysRevC.100.015805}. The lower one at E$_p$\,=\,278\,keV is especially important as it serves as a normalization point in several experiments. Its strength was found to be in excellent agreement with the available data. The strength of the E$_p$\,=\,1058\,keV resonance, on the other hand, was measured to be significantly higher than the previous data. This also indicates that further measurements of this important reaction is necessary. Figure \ref{fig:strengths} shows the strengths of the two studied resonances from the literature and from the present work. 

\subsection{Experiments related to the astrophysical p-process}

Nuclear reactions that take place in stellar hydrogen burning processes were typically investigated experimentally several times in the past. As it was emphasized above, it is the precision and reliability of the data which is needed to be increased. In heavy element nucleosynthesis processes, on the other hand, nuclear reactions are involved that were never studied experimentally. A good example is the astrophysical p-process \cite{Rauscher_2013} which involves thousands of reactions. In lack of experimental data, the reaction rates are obtained from theoretical cross section calculations which exhibit high uncertainty. The experimental study of p-process related reactions is therefore highly needed in order to test the calculations and to provide reliable input parameters for the models.

Owing to the low cross sections, the Gamow window typically cannot be covered by experimental data, especially in the case of reactions involving alpha particles. Therefore, measurement at higher energies and in wide energy ranges are again necessary as some of the input parameters of the theoretical calculations can be checked at least at higher energies. For example, the most uncertain input parameter for alpha-induced reactions is the alpha-nucleus optical potential. The study of this quantity at higher energies may help constrain the reaction cross sections in the Gamow window as well.

In recent years our systematic study of the alpha-nucleus optical potential \cite{5f994521ea7c4fdc97aea9f787fb4b97} has been extended to the completely unexplored mass region of the heaviest p-isotopes. Alpha-induced cross section data on $^{191,193}$Ir and $^{197}$Au are now available. For details see the original publications: \cite{SZUCS2018396,PhysRevC.100.065803}. The results indicate the need for developing new or modified optical potentials for low energy problems \cite{MOHR2013651} and based on the observations a new treatment of the alpha-induced reactions has also been suggested \cite{mohr_thisvolume}.

\section{Summary}

As astronomical observations as well as astrophysical models are becoming more and more precise, the knowledge about nuclear reactions of astrophysical relevance must also be improved. In many cases, experiments carried out at energies higher than the astrophysically relevant ones can be very important in reaching this goal. In this paper a few examples were briefly described emphasizing the importance of precise experimental data in wide energy ranges. 

\subsection*{Acknowledgments}

The author thanks the members of the Atomki nuclear astrophysics group who carried out the research presented in this paper. This work was supported by NKFIH (grants K120666 and NN128072) and by the European Cooperation in Science and Technology (”ChETEC” COST Action, CA16117).

\section*{References}
\providecommand{\newblock}{}


\begin{thebibliography}{10}
\expandafter\ifx\csname url\endcsname\relax
  \def\url#1{{\tt #1}}\fi
\expandafter\ifx\csname urlprefix\endcsname\relax\def\urlprefix{URL }\fi
\providecommand{\eprint}[2][]{\url{#2}}

\bibitem{BROGGINI201855}
Broggini C, Bemmerer D, Caciolli A and Trezzi D 2018 {\em Progress in Particle
  and Nuclear Physics\/} {\bf 98} 55 -- 84 ISSN 0146-6410
  \urlprefix\url{http://www.sciencedirect.com/science/article/pii/S0146641017300868}

\bibitem{Gyurky2019}
Gy{\"u}rky G, F{\"u}l{\"o}p Z, K{\"a}ppeler F, Kiss G~G and Wallner A 2019 {\em
  The European Physical Journal A\/} {\bf 55} 41 ISSN 1434-601X
  \urlprefix\url{https://doi.org/10.1140/epja/i2019-12708-4}

\bibitem{PhysRevLett.109.202501}
Scott D~A, Caciolli A, Di~Leva A, Formicola A, Aliotta M, Anders M, Bemmerer D,
  Broggini C, Campeggio M, Corvisiero P, Elekes Z, F\"ul\"op Z, Gervino G,
  Guglielmetti A, Gustavino C, Gy\"urky G, Imbriani G, Junker M, Laubenstein M,
  Menegazzo R, Marta M, Napolitani E, Prati P, Rigato V, Roca V, Somorjai E,
  Salvo C, Straniero O, Strieder F, Sz\"ucs T, Terrasi F and Trezzi D (LUNA
  Collaboration) 2012 {\em Phys. Rev. Lett.\/} {\bf 109}(20) 202501
  \urlprefix\url{https://link.aps.org/doi/10.1103/PhysRevLett.109.202501}

\bibitem{ROLFS197329}
Rolfs C 1973 {\em Nuclear Physics A\/} {\bf 217} 29 -- 70 ISSN 0375-9474
  \urlprefix\url{http://www.sciencedirect.com/science/article/pii/0375947473906222}

\bibitem{PhysRevC.86.055801}
Kontos A, G\"orres J, Best A, Couder M, deBoer R, Imbriani G, Li Q, Robertson
  D, Sch\"urmann D, Stech E, Uberseder E and Wiescher M 2012 {\em Phys. Rev.
  C\/} {\bf 86}(5) 055801
  \urlprefix\url{https://link.aps.org/doi/10.1103/PhysRevC.86.055801}

\bibitem{PhysRevC.95.035805}
Gy\"urky G, Ornelas A, F\"ul\"op Z, Hal\'asz Z, Kiss G~G, Sz\"ucs T, Husz\'ank
  R, Horny\'ak I, Rajta I and Vajda I 2017 {\em Phys. Rev. C\/} {\bf 95}(3)
  035805 \urlprefix\url{https://link.aps.org/doi/10.1103/PhysRevC.95.035805}

\bibitem{BORDEANU20131}
Bordeanu C, Gy{\"u}rky G, Hal{\'a}sz Z, Sz{\"u}cs T, Kiss G, Elekes Z, Farkas
  J, F{\"u}l{\"o}p Z and Somorjai E 2013 {\em Nuclear Physics A\/} {\bf 908} 1
  -- 11 ISSN 0375-9474
  \urlprefix\url{http://www.sciencedirect.com/science/article/pii/S0375947413003606}

\bibitem{HE2013287}
He J, Chen S, Rolfs C, Xu S, Hu J, Ma X, Wiescher M, deBoer R, Kajino T,
  Kusakabe M, Zhang L, Hou S, Yu X, Zhang N, Lian G, Zhang Y, Zhou X, Xu H,
  Xiao G and Zhan W 2013 {\em Physics Letters B\/} {\bf 725} 287 -- 291 ISSN
  0370-2693
  \urlprefix\url{http://www.sciencedirect.com/science/article/pii/S0370269313006084}

\bibitem{PhysRevC.99.055804}
Sz\"ucs T, Kiss G~G, Gy\"urky G, Hal\'asz Z, Szegedi T~N and F\"ul\"op Z 2019
  {\em Phys. Rev. C\/} {\bf 99}(5) 055804
  \urlprefix\url{https://link.aps.org/doi/10.1103/PhysRevC.99.055804}

\bibitem{doi:10.1146/annurev.nucl.012809.104505}
Wiescher M, G{\"o}rres J, Uberseder E, Imbriani G and Pignatari M 2010 {\em
  Annual Review of Nuclear and Particle Science\/} {\bf 60} 381--404
  (\textit{Preprint}
  \eprint{https://doi.org/10.1146/annurev.nucl.012809.104505})
  \urlprefix\url{https://doi.org/10.1146/annurev.nucl.012809.104505}

\bibitem{PhysRevC.100.015805}
Gy\"urky G, Hal\'asz Z, Kiss G~G, Sz\"ucs T, Cs\'{\i}k A, T\"or\"ok Z,
  Husz\'ank R, Kohan M~G, Wagner L and F\"ul\"op Z 2019 {\em Phys. Rev. C\/}
  {\bf 100}(1) 015805
  \urlprefix\url{https://link.aps.org/doi/10.1103/PhysRevC.100.015805}

\bibitem{Rauscher_2013}
Rauscher T, Dauphas N, Dillmann I, Fr{\"o}hlich C, F{\"u}l{\"o}p Z and
  Gy{\"u}rky G 2013 {\em Reports on Progress in Physics\/} {\bf 76} 066201

\bibitem{5f994521ea7c4fdc97aea9f787fb4b97}
Gy{\"u}rky G, Hal{\'a}sz Z, Sz{\"u}cs T, Kiss G and F{\"u}l{\"o}p Z 2016 {\em
  Journal of Physics: Conference Series\/} {\bf 665} ISSN 1742-6588

\bibitem{SZUCS2018396}
Sz{\"u}cs T, Kiss G, Gy{\"u}rky G, Hal{\'a}ász Z, F{\"u}l{\"o}p Z and Rauscher
  T 2018 {\em Physics Letters B\/} {\bf 776} 396 -- 401 ISSN 0370-2693
  \urlprefix\url{http://www.sciencedirect.com/science/article/pii/S0370269317309644}

\bibitem{PhysRevC.100.065803}
Sz\"ucs T, Mohr P, Gy\"urky G, Hal\'asz Z, Husz\'ank R, Kiss G~G, Szegedi T~N,
  T\"or\"ok Z and F\"ul\"op Z 2019 {\em Phys. Rev. C\/} {\bf 100}(6) 065803
  \urlprefix\url{https://link.aps.org/doi/10.1103/PhysRevC.100.065803}

\bibitem{MOHR2013651}
Mohr P, Kiss G, F{\"u}l{\"o}p Z, Galaviz D, Gy{\"u}rky G and Somorjai E 2013
  {\em Atomic Data and Nuclear Data Tables\/} {\bf 99} 651 -- 679 ISSN
  0092-640X
  \urlprefix\url{http://www.sciencedirect.com/science/article/pii/S0092640X13000545}

\bibitem{mohr_thisvolume}
Mohr P \it, in this volume

\end{thebibliography}
\end{document}